\documentclass[preprint]{aastex}
\begin{document}
\title{VERITAS OBSERVATIONS OF DAY-SCALE FLARING OF M\,87 IN 2010 APRIL}
\author{
E.~Aliu\altaffilmark{1},
T.~Arlen\altaffilmark{2},
T.~Aune\altaffilmark{3},
M.~Beilicke\altaffilmark{4},
W.~Benbow\altaffilmark{5},
A.~Bouvier\altaffilmark{3},
S.~M.~Bradbury\altaffilmark{6},
J.~H.~Buckley\altaffilmark{4},
V.~Bugaev\altaffilmark{4},
K.~Byrum\altaffilmark{7},
A.~Cannon\altaffilmark{8},
A.~Cesarini\altaffilmark{9},
L.~Ciupik\altaffilmark{10},
E.~Collins-Hughes\altaffilmark{8},
M.~P.~Connolly\altaffilmark{9},
W.~Cui\altaffilmark{11},
R.~Dickherber\altaffilmark{4},
C.~Duke\altaffilmark{12},
M.~Errando\altaffilmark{1},
A.~Falcone\altaffilmark{13},
J.~P.~Finley\altaffilmark{11},
G.~Finnegan\altaffilmark{14},
L.~Fortson\altaffilmark{15},
A.~Furniss\altaffilmark{3},
N.~Galante\altaffilmark{5},
D.~Gall\altaffilmark{16},
S.~Godambe\altaffilmark{14},
S.~Griffin\altaffilmark{17},
J.~Grube\altaffilmark{10},
R.~Guenette\altaffilmark{17},
G.~Gyuk\altaffilmark{10},
D.~Hanna\altaffilmark{17},
J.~Holder\altaffilmark{18},
H.~Huan\altaffilmark{19},
G.~Hughes\altaffilmark{20},
C.~M.~Hui\altaffilmark{14},
T.~B.~Humensky\altaffilmark{21},
A.~Imran\altaffilmark{22},
P.~Kaaret\altaffilmark{16},
N.~Karlsson\altaffilmark{15},
M.~Kertzman\altaffilmark{23},
D.~Kieda\altaffilmark{14},
H.~Krawczynski\altaffilmark{4},
F.~Krennrich\altaffilmark{22},
M.~J.~Lang\altaffilmark{9},
S.~LeBohec\altaffilmark{14},
A.~S~Madhavan\altaffilmark{22},
G.~Maier\altaffilmark{20},
P.~Majumdar\altaffilmark{2},
S.~McArthur\altaffilmark{4},
A.~McCann\altaffilmark{17},
P.~Moriarty\altaffilmark{24},
R.~Mukherjee\altaffilmark{1},
P.~D~Nu\~{n}ez\altaffilmark{14},
R.~A.~Ong\altaffilmark{2},
M.~Orr\altaffilmark{22},
A.~N.~Otte\altaffilmark{25},
N.~Park\altaffilmark{19},
J.~S.~Perkins\altaffilmark{26,27},
A.~Pichel\altaffilmark{28},
M.~Pohl\altaffilmark{20,29},
H.~Prokoph\altaffilmark{20},
J.~Quinn\altaffilmark{8},
K.~Ragan\altaffilmark{17},
L.~C.~Reyes\altaffilmark{30},
P.~T.~Reynolds\altaffilmark{31},
E.~Roache\altaffilmark{5},
H.~J.~Rose\altaffilmark{6},
J.~Ruppel\altaffilmark{20,29},
D.~B.~Saxon\altaffilmark{18},
M.~Schroedter\altaffilmark{5},
G.~H.~Sembroski\altaffilmark{11},
G.~D.~\c{S}ent\"{u}rk\altaffilmark{21},
C.~Skole\altaffilmark{20},
D.~Staszak\altaffilmark{17},
G.~Te\v{s}i\'{c}\altaffilmark{17},
M.~Theiling\altaffilmark{11},
S.~Thibadeau\altaffilmark{4},
K.~Tsurusaki\altaffilmark{16},
J.~Tyler\altaffilmark{17},
A.~Varlotta\altaffilmark{11},
V.~V.~Vassiliev\altaffilmark{2},
S.~Vincent\altaffilmark{14},
M.~Vivier\altaffilmark{18},
S.~P.~Wakely\altaffilmark{19},
J.~E.~Ward\altaffilmark{8},
T.~C.~Weekes\altaffilmark{5},
A.~Weinstein\altaffilmark{22},
T.~Weisgarber\altaffilmark{19},
D.~A.~Williams\altaffilmark{3}, 
B.~Zitzer\altaffilmark{11}
}

\altaffiltext{1}{Department of Physics and Astronomy, Barnard College, Columbia University, NY 10027, USA}
\altaffiltext{2}{Department of Physics and Astronomy, University of California, Los Angeles, CA 90095, USA}
\altaffiltext{3}{Santa Cruz Institute for Particle Physics and Department of Physics, University of California, Santa Cruz, CA 95064, USA}
\altaffiltext{4}{Department of Physics, Washington University, St. Louis, MO 63130, USA}
\altaffiltext{5}{Fred Lawrence Whipple Observatory, Harvard-Smithsonian Center for Astrophysics, Amado, AZ 85645, USA}
\altaffiltext{6}{School of Physics and Astronomy, University of Leeds, Leeds, LS2 9JT, UK}
\altaffiltext{7}{Argonne National Laboratory, 9700 S. Cass Avenue, Argonne, IL 60439, USA}
\altaffiltext{8}{School of Physics, University College Dublin, Belfield, Dublin 4, Ireland}
\altaffiltext{9}{School of Physics, National University of Ireland Galway, University Road, Galway, Ireland}
\altaffiltext{10}{Astronomy Department, Adler Planetarium and Astronomy Museum, Chicago, IL 60605, USA}
\altaffiltext{11}{Department of Physics, Purdue University, West Lafayette, IN 47907, USA}
\altaffiltext{12}{Department of Physics, Grinnell College, Grinnell, IA 50112-1690, USA}
\altaffiltext{13}{Department of Astronomy and Astrophysics, 525 Davey Lab, Pennsylvania State University, University Park, PA 16802, USA}
\altaffiltext{14}{Department of Physics and Astronomy, University of Utah, Salt Lake City, UT 84112, USA; cmhui@physics.utah.edu}
\altaffiltext{15}{School of Physics and Astronomy, University of Minnesota, Minneapolis, MN 55455, USA}
\altaffiltext{16}{Department of Physics and Astronomy, University of Iowa, Van Allen Hall, Iowa City, IA 52242, USA}
\altaffiltext{17}{Physics Department, McGill University, Montreal, QC H3A 2T8, Canada}
\altaffiltext{18}{Department of Physics and Astronomy and the Bartol Research Institute, University of Delaware, Newark, DE 19716, USA}
\altaffiltext{19}{Enrico Fermi Institute, University of Chicago, Chicago, IL 60637, USA}
\altaffiltext{20}{DESY, Platanenallee 6, 15738 Zeuthen, Germany}
\altaffiltext{21}{Physics Department, Columbia University, New York, NY 10027, USA}
\altaffiltext{22}{Department of Physics and Astronomy, Iowa State University, Ames, IA 50011, USA}
\altaffiltext{23}{Department of Physics and Astronomy, DePauw University, Greencastle, IN 46135-0037, USA}
\altaffiltext{24}{Department of Life and Physical Sciences, Galway-Mayo Institute of Technology, Dublin Road, Galway, Ireland}
\altaffiltext{25}{School of Physics \& Center for Relativistic Astrophysics, Georgia Institute of Technology, 837 State Street NW, Atlanta, GA 30332-0430}
\altaffiltext{26}{CRESST and Astroparticle Physics Laboratory NASA/GSFC, Greenbelt, MD 20771, USA.}
\altaffiltext{27}{University of Maryland, Baltimore County, 1000 Hilltop Circle, Baltimore, MD 21250, USA.}
\altaffiltext{28}{Instituto de Astronomia y Fisica del Espacio, Casilla de Correo 67-Sucursal 28, (C1428ZAA) Ciudad Autónoma de Buenos Aires, Argentina}
\altaffiltext{29}{Institut f\"ur Physik und Astronomie, Universit\"at Potsdam, 14476 Potsdam-Golm,Germany}
\altaffiltext{30}{Physics Department, California Polytechnic State University, San Luis Obispo, CA 94307, USA}
\altaffiltext{31}{Department of Applied Physics and Instrumentation, Cork Institute of Technology, Bishopstown, Cork, Ireland}

\begin{abstract}
VERITAS has been monitoring the very-high-energy (VHE; $>$\,100\,GeV) gamma-ray
activity of the radio galaxy M\,87 since 2007.  During 2008, flaring
activity on a timescale of a few days was observed with a peak flux of $(0.70 \pm 0.16) \times 10^{-11}\,\mathrm{cm}^{-2}\,\mathrm{s}^{-1}$ at energies above 350\,GeV.  In 2010 April, VERITAS detected a flare from
M\,87 with peak flux of $(2.71 \pm 0.68) \times 10^{-11}\,\mathrm{cm}^{-2}\,\mathrm{s}^{-1}$ for $E\,>$\,350\,GeV.  The
source was observed for six consecutive nights during the flare, resulting in a
total of 21 hr of good quality data.  The most rapid flux variation occurred on the trailing edge of
the flare with an exponential flux decay time of $0.90^{ +0.22}_{
  -0.15}$\,days.  The shortest detected exponential rise time is three
times as long, at $2.87^{ +1.65}_{-0.99}$\,days.  The quality of the
data sample is such
that spectral analysis can be performed for three periods: rising
flux, peak flux, and falling flux.  The spectra obtained are
consistent with power-law forms.  The spectral index at the 
peak of the flare is equal to $2.19 \pm 0.07$. There is some
indication that the spectrum is softer in the falling
phase of the flare than the peak phase, with a confidence level corresponding to 3.6 standard deviations.  We
discuss the implications of these results for the acceleration and
cooling rates of VHE electrons in M\,87 and the constraints they
provide on the physical size of the emitting region. 
\end{abstract}

\section{Introduction}
M\,87 is a giant radio galaxy located in the Virgo cluster at a distance
of 16.7\,Mpc \citep{mei07}. It is believed to harbor a supermassive black hole of mass
$(3.2 \pm 0.9) \times 10^9 M_{\odot}$ \citep{macchetto97}, derived from gas kinematics, or
 $(6.6 \pm 0.4) \times 10^9 M_{\odot}$ \citep{gebhardt11}, derived from stellar kinematics.  Its jet is misaligned with the line of sight; this, along with the proximity of M\,87, allows for
detailed observations of its structure in the radio \citep[e.g.,][]{cheung07},
optical \citep[e.g.,][]{biretta99}, and X-ray \citep[e.g.,][]{marshall02,wilson02}
wavebands.  Apparent superluminal
motion is observed in the radio and optical wavebands
\citep{biretta95, biretta99}.  Month-scale flaring activity
has been observed in various energy ranges at the nucleus and at
HST-1, the jet feature closest to the nucleus
\citep{perlman03,harris09}.  The jet knot HST-1 is located 0.85
arcsec ($\approx$ 69\,pc projected) from the nucleus and is
resolved from the nucleus in the radio, optical, and X-ray energy bands.

Very-high-energy (VHE) gamma-ray emission from M\,87 was first detected
by HEGRA in 1998/1999 at energies above 730\,GeV \citep{hegra03} and
has since been confirmed by H.E.S.S. \citep{hess06}, VERITAS
\citep{veritas08}, and MAGIC \citep{magic08}.  The first gamma-ray flaring
activity from M\,87 was reported by H.E.S.S. in 2005, with the flux varying on a timescale of days.  The angular resolution of current
ground-based gamma-ray instruments is not sufficient to distinguish
the different morphological features in M\,87, which is therefore
detected as a point-like source in VHE gamma rays.  However, given the
short timescale of the flare, the characteristic size of the
gamma-ray emitting region (or of moving regions of
low gamma-ray opacity between us and the emitting region) is constrained by the light crossing
time of these features and the relativistic Doppler factor of their
motion in the observer's reference frame.  Under the size constraint,
the two most likely regions for gamma-ray production are the
unresolved nucleus and the HST-1 knot \citep{hess06,cheung07}.

During the 2005 gamma-ray flare observed by H.E.S.S., \emph{Chandra}
reported historically maximal flaring from HST-1
\citep{harris06}.  Through the causality argument, the timescale of
the enhanced TeV emission implies an emission region
size of about $R \le 5 \times 10^{15}\,\delta$\,cm, where
$\delta$ is the Doppler factor of the radiating region.
\citet{hess06} preferred the nucleus over HST-1 as the VHE gamma-ray
production region due to an unrealistically small opening angle ($\sim 1.5
\times 10^{-3}\,\delta$\,deg) required to channel energy from the
central object to the HST-1 knot.  However,
the Very Long Baseline Array (VLBA) imaged compact knots in HST-1 that are not resolved with semi-minor axes $\le 5 \times 10^{17}$\,cm
\citep{cheung07}, and \citet{stawarz06} proposed jet reconfinement
at the HST-1 location which can in turn produce TeV emission.
Therefore, HST-1 remains a candidate for TeV emission.  However, a VHE gamma-ray flare in 2008 coincided with the historical maximal X-ray
flux from the nucleus detected by \emph{Chandra}, while HST-1 remained
in a low state at that time and its X-ray flux was below that of the nucleus.  In
addition, increasing radio flux from the nucleus, but not from the
jet, was observed by the VLBA, lasting up to two months past the VHE
gamma-ray flare \citep{joint}.  The 2008 observations therefore favor the nucleus as the origin of the VHE gamma-ray emission.

After the launch of the \textit{Fermi Gamma-ray Space Telescope} in
the summer of 2008, M\,87 was also detected in the MeV--GeV energy range
by the \textit{Fermi} Large Area Telescope \citep[LAT;][]{fermi09}.
However, no significant flaring activity was detected in 2009 at any
wavelength. 

M\,87 has been monitored every year in VHE gamma rays since 2003 by at
least one of the three major atmospheric-Cherenkov telescope arrays ---
H.E.S.S., MAGIC, and VERITAS.  In 2010, VHE flaring activity up to 20\% of
the Crab Nebula flux was detected from M\,87 in the span of several
days \citep{atel2542}, and gamma-ray, X-ray, optical, and radio observations were subsequently triggered.  Detailed results from the VERITAS
observations are presented in this paper, and the multiwavelength
light curve will be presented in a separate publication \citep{joint2010}.

\section{Observations and Analysis}
VERITAS is an array of four 12\,m diameter imaging atmospheric
Cherenkov telescopes located at the Fred Lawrence Whipple Observatory
in southern Arizona, 1.3\,km above sea level.  The telescopes are 
situated approximately 100\,m apart, forming a convex quadrilateral.  Each
telescope is equipped with a camera of 499 photomultiplier tubes (PMTs)
arranged in a hexagonal lattice covering a field of view with a
diameter of $3.5^\circ$.  The array is sensitive to photons with
energy from $\sim 150$\,GeV to more than 30\,TeV, with 
an angular resolution of $\sim 0.1^\circ$ and an effective area
of $\sim 10^5\,\mathrm{m}^2$ at 1\,TeV.  Further description of the
VERITAS observatory and its performance are given in \citet{Perkins09}
and \citet{holder06}.

M\,87 was observed between 2009 December and 2010 May for 53.1 hr.
Observations were conducted at a range of zenith angles between
$19^\circ$ and $40^\circ$, with low elevation excursions (up to
$60^\circ$ from zenith) during the nights of April 9 through 11 when episodes
of flaring were detected.  More than 95\% of the data were taken with
the full four-telescope array and the remainder with a three-telescope
sub-array.  To enable simultaneous estimation of source and background
signals, the data were accumulated in ``wobble mode'' for which the
source is offset from the camera center by $0.5^\circ$ in alternating
directions every 20 minutes.  The analysis presented in this paper is
based on 44.6 hr of live time which satisfied data quality and integrity selection criteria.   

The data are analyzed with the algorithm described in \citet{veri10}.
Atmospheric gamma-ray shower images are first corrected for dispersion
in PMT gain and timing using information obtained from nightly laser
calibrations \citep{hanna07}.  Then, an image-cleaning process is
applied to select pixels with a signal significantly above the
night-sky background level.  After cleaning, the images are
parameterized \citep{hillas85} and the shower direction is
reconstructed using the stereoscopic technique \citep{hofmann99}.
Events are then selected as gamma-ray-like if at least three camera
images pass selection criteria optimized for a source with 1\% of the Crab Nebula flux.  The
results reported in this paper have all been confirmed by an
independent secondary analysis package described in \citet{daniel07}.

\section{Results}
During the six-month observation period, M\,87 was detected at a level of
25.6\,standard deviation ($\sigma$) above the background, with an
average flux of $(5.44 \pm 0.30) \times 10^{-12}\,\mathrm{photon}\,
\mathrm{cm}^{-2}\,\mathrm{s}^{-1}$ at energies above 350\,GeV,
equivalent to 5\% of the Crab Nebula flux above 350\,GeV.  The following sub-sections
first present the daily light curve obtained over six months of
observation, then the April flaring episode light curve binned in
20 minute intervals, followed by the timescale and spectral analyses
of the April flare.

\subsection{Daily Flux Over a Six-month Period}
\begin{figure}
\plotone{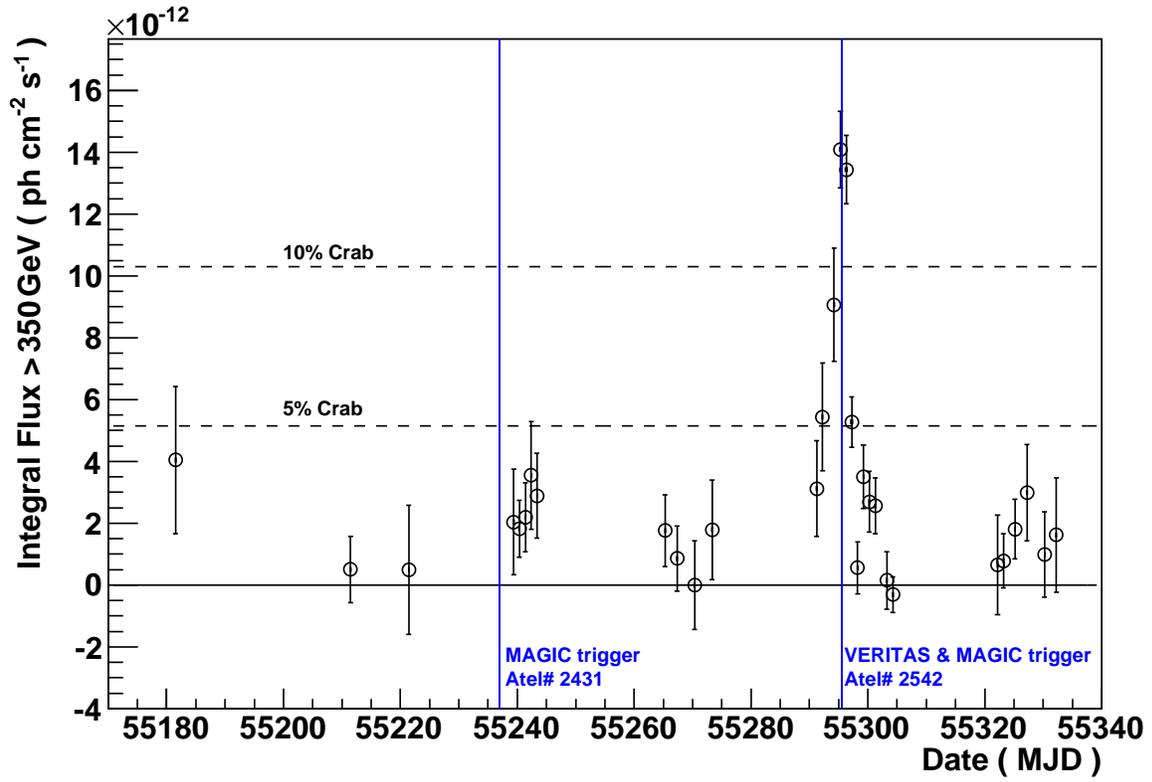}
\caption{Daily light curve of M\,87 observed by VERITAS in 2010.  Clear evidence of flaring activity is seen in 2010 April (MJD 55291--55298).  Trigger alerts sent by MAGIC on February 10 (MJD 55237) and by VERITAS and MAGIC on April 9 (MJD 55295) are indicated by vertical lines.  The average nightly flux during the peak of the flare exceeds 10\% of the Crab Nebula flux at the same energy threshold of 350\,GeV.  A constant spectral index of 2.5 was assumed for the daily flux calculation.}
\label{overallLC}
\end{figure}

Figure \ref{overallLC} shows the daily flux recorded by VERITAS
between 2009 December and 2010 May.  Applying a constant-flux
fit to the daily light curve gives a $\chi^2$/dof value of 269.4/29, a strong
indication the flux was not constant during the observation
period.  

In 2010 February, the MAGIC Collaboration reported an increased
activity of M\,87 with more than 10\% of the Crab Nebula flux on February 9
\citep{magicatel}.  At that time, VERITAS observations were hampered by 
poor weather conditions, but M\,87 was detected by VERITAS in a typical state two nights after the MAGIC alert.  In 2010 April, VERITAS detected M\,87 with
an elevated flux during a week of observations between April 5 and
11 and triggered subsequent multiwavelength observations
\citep{atel2542}. 

\subsection{The 2010 April Flare}
\begin{figure}
\plotone{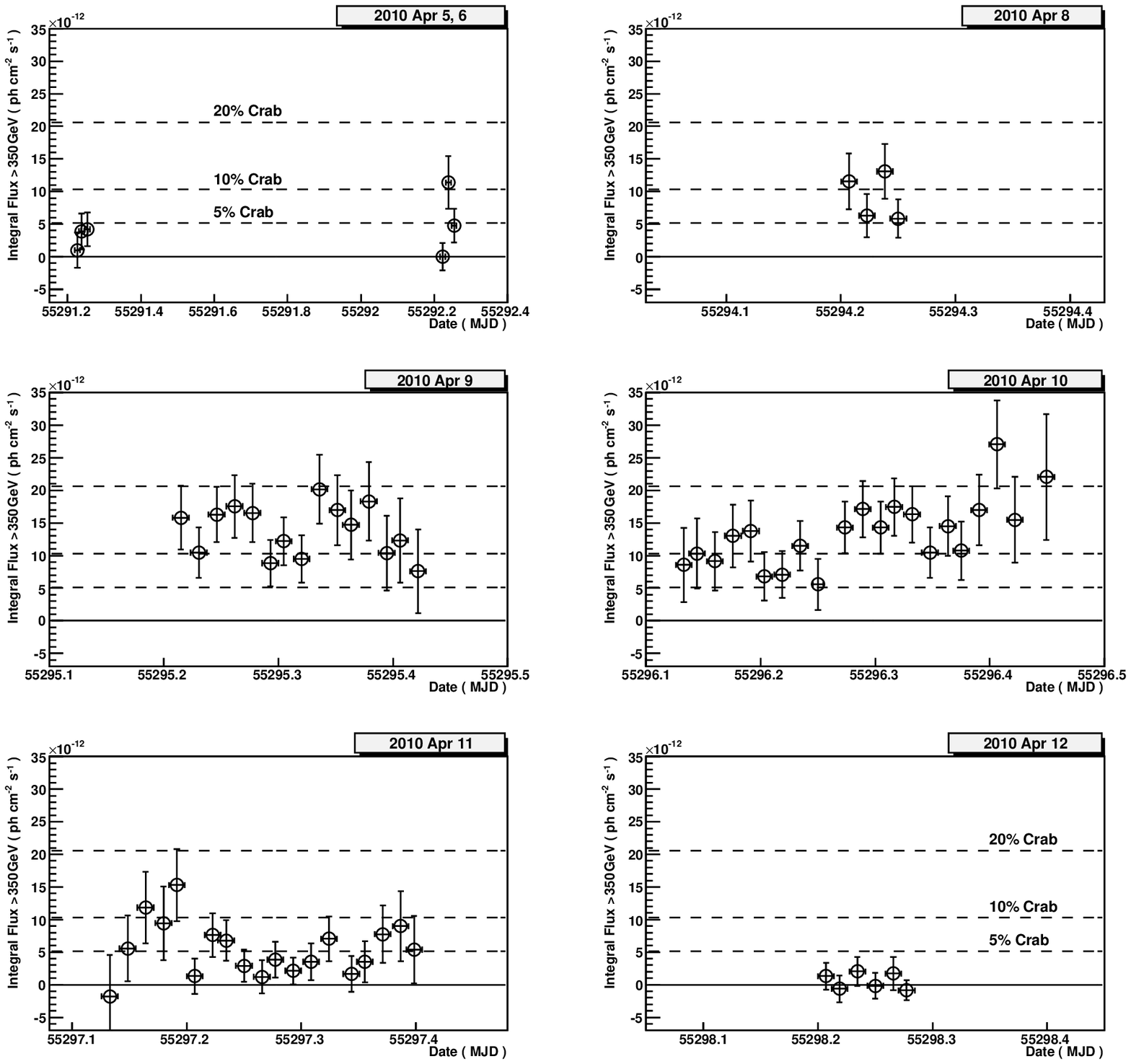}
\caption{VERITAS light curve with 20 minute binning during the
  flare period between 2010  April 5 and April 12 (MJD 55291--55298).  The
  flux scale is the same for all six panels, and dashed lines indicating
  5\%, 10\%, and 20\% of the Crab Nebula flux are included.  A
  constant spectral index of 2.5 was assumed for the flux calculation.}
\label{flareLC}
\end{figure}

Figure \ref{flareLC} shows the light curve binned in 20 minute
intervals during the flare.  Observations around the peak of the flare
were carried out up to high zenith angles; this reduces the
sensitivity at low energies, and as a result, the data points taken at the
end of April 10 and the beginning of April 11 have larger uncertainties.  The flaring episode began with increasing flux during the nights of April 5 and 6, reaching 10\% of the Crab
Nebula flux on April 8.  On the following three nights, M\,87 was
observed for more than five hours each night.  The average flux on
April 9 and 10 was 15\% of the Crab Nebula flux, reaching as much as 20\% of the Crab Nebula flux in individual 20 minute bins.  The average flux on April 11th
was 5\% of the Crab Nebula flux.  VERITAS continued to monitor M\,87 for two hours each night
from April 12 to 15, when the flux level returned to a few percent
of the Crab Nebula flux, comparable to the low-state flux
measured in the past.  All flux comparison with the Crab Nebula is at energies above 350\,GeV.

\subsubsection{Flux Variability Timescale Analysis}
Using the data from April 9 and 10 (MJD 55295 and 55296) when
maximal activity occurred, we searched for variability
within each night.  On April 9, fifteen 20 minute exposures were
taken in total and a constant-flux fit yields a $\chi^2$/dof value of
9.3/14 and a corresponding $\chi^2$ probability of 0.81.  On April
10th, twenty-one 20-minute exposures were taken and the 
constant-flux fit gives a $\chi^2$/dof value of 19.8/20 and a corresponding
$\chi^2$ probability of 0.47.  In order to investigate variability
within a single day of observation in more detail, the wavelet
analysis described by \citet{wavelet} is applied to the April 9
and 10 data sets.  The highest confidence level for the April 9
data set is obtained for a variability timescale of 80 minutes.
However, the confidence level is only 86.2\%, implying that on an
80 minute timescale, the evidence for variability is only at the
level of $1.5\,\sigma$.  The highest confidence level for the April
10 data set is obtained for a variability timescale of 160 minutes
at 97.5\%, or $2.2\,\sigma$.  Therefore, no evidence for intra-night
variability is found. 

\begin{figure}
\plotone{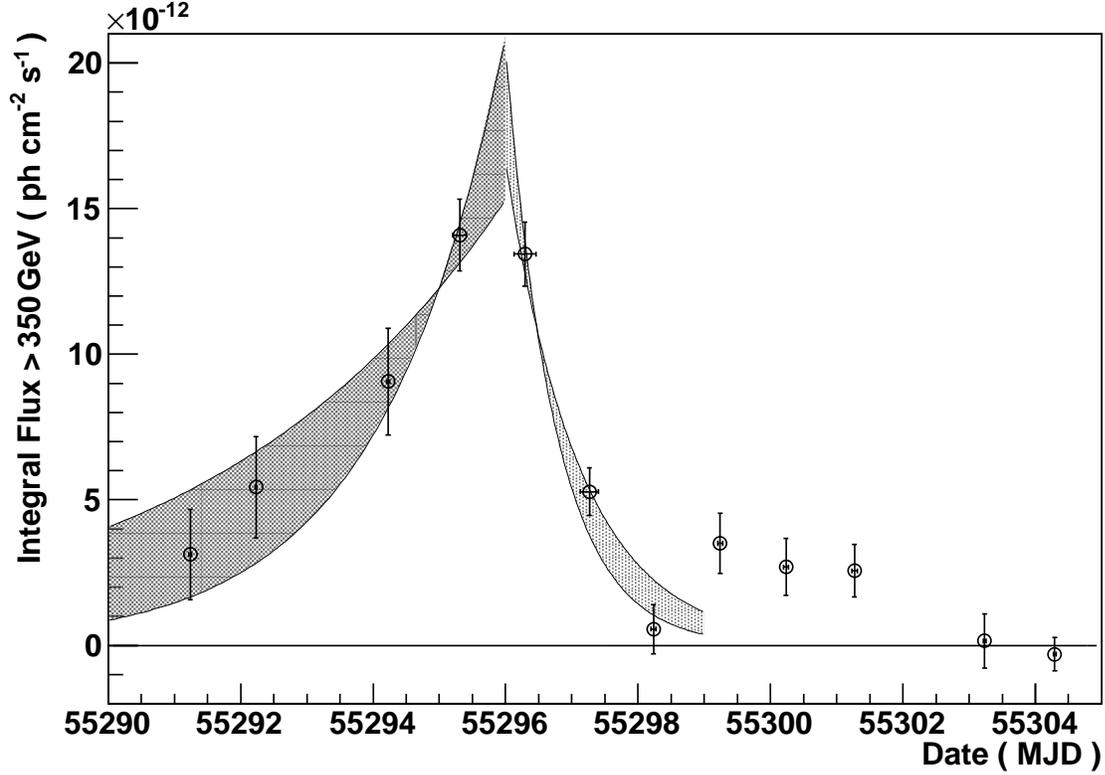}
\caption{Fits to the 2010 April VHE gamma-ray light curve of M\,87 leading up to the
  flare and trailing the flare, with fit errors included and shown as shaded regions.  The exponential timescale is
  $2.87^{+1.65}_{-0.99}$\,days for the rising flux portion, and
  $0.90^{+0.22}_{-0.15}$\,days for MJD 55296--55298 segment of the
  falling flux.} 
\label{lightexp}
\end{figure}

Figure \ref{lightexp} shows the daily light curve of the April
flaring episode.  To characterize the timescales of the flare, an
exponential function of the form $\Phi = p_0\,e^{(t-55290)/p_1}$ is fitted to
different periods of the April daily light curve by $\chi^2$
minimization.  The parameter $p_1$ represents the characteristic time
of the flux variation.  For the days leading up to the flare (MJD 55291--55295, April 5--9), the 
minimal $\chi^2$/dof value of 0.3/2 is obtained for $p_1 = 2.87$\,days.
The error bars of the fit parameters $p_0$ and $p_1$ are
determined by finding the parameter ranges with $\chi^2$ between
$\chi^2_{\mathrm{min}}$ and $\chi^2_{\mathrm{min}} + 2.30$, where
$\chi^2_{\mathrm{min}}$ is the smallest $\chi^2$ value.  The same
$\chi^2$ calculation 
is repeated for data from the peak flux onward.  The details are
presented in Table \ref{lc-fittab}.  For the period between MJD 55296
and 55304 (April 10--18), the exponential decay time is $1.12^{+0.31}_{-0.26}$\,days.  An even shorter decay time of
$0.90^{+0.22}_{-0.15}$\,days is obtained by restricting the fit to the
period between MJD 55296 and 55298 (April 10--12).  To
investigate the possibility of a second flare between MJD 55299 and
55301, a constant-flux fit is applied to data points between MJD
55298 and 55304 (April 12--18).  The $\chi^2$/dof value of the
constant-flux fit is 9.6/4 with a corresponding $\chi^2$ probability
of 0.05.  In spite of this low confidence level for the constant-flux
hypothesis, there is nevertheless insufficient evidence to confirm the
presence of a second, separated flare component around MJD 55299--55301 (April 13--15). 

\begin{deluxetable}{cccccc}
\tablewidth{0pt}
\tablecaption{$\chi^2$ minimized parameters of the April flare light curve
  in Figure \ref{lightexp} (Fit function $\Phi = p_0\,e^{(t-55290)/p_1}$).  The error bars of $p_0$ and $p_1$ are statistical only.}
\tablehead{\colhead{Period (MJD)} & \colhead{$\chi^2$/dof} &
  \colhead{$\chi^2$ Probability} & \colhead{$p_0$ ($\rm{cm}^{-2}\rm{\,s}^{-1}$)} & \colhead{$p_1$
    (days)} \label{lc-fittab} }
\startdata
55291--55295 & 0.3/2 & 0.88 & $2.20^{+1.86}_{-1.34} \times 10^{-12}$ & $2.87^{+1.65}_{-0.99}$ \\
55296--55304 & 23.7/6 & $6.0 \times 10^{-4}$ & $3.61^{+17.64}_{-2.61} \times 10^{-9}$ & $-(1.12^{+0.31}_{-0.26})$ \\ 
55296--55298 & 2.1/1 & 0.15 & $1.48^{+4.62}_{-1.32} \times 10^{-8}$ & $-(0.90^{+0.22}_{-0.15})$ \\
\enddata
\end{deluxetable}

\subsubsection{Spectral Analysis}
\begin{figure}
\plotone{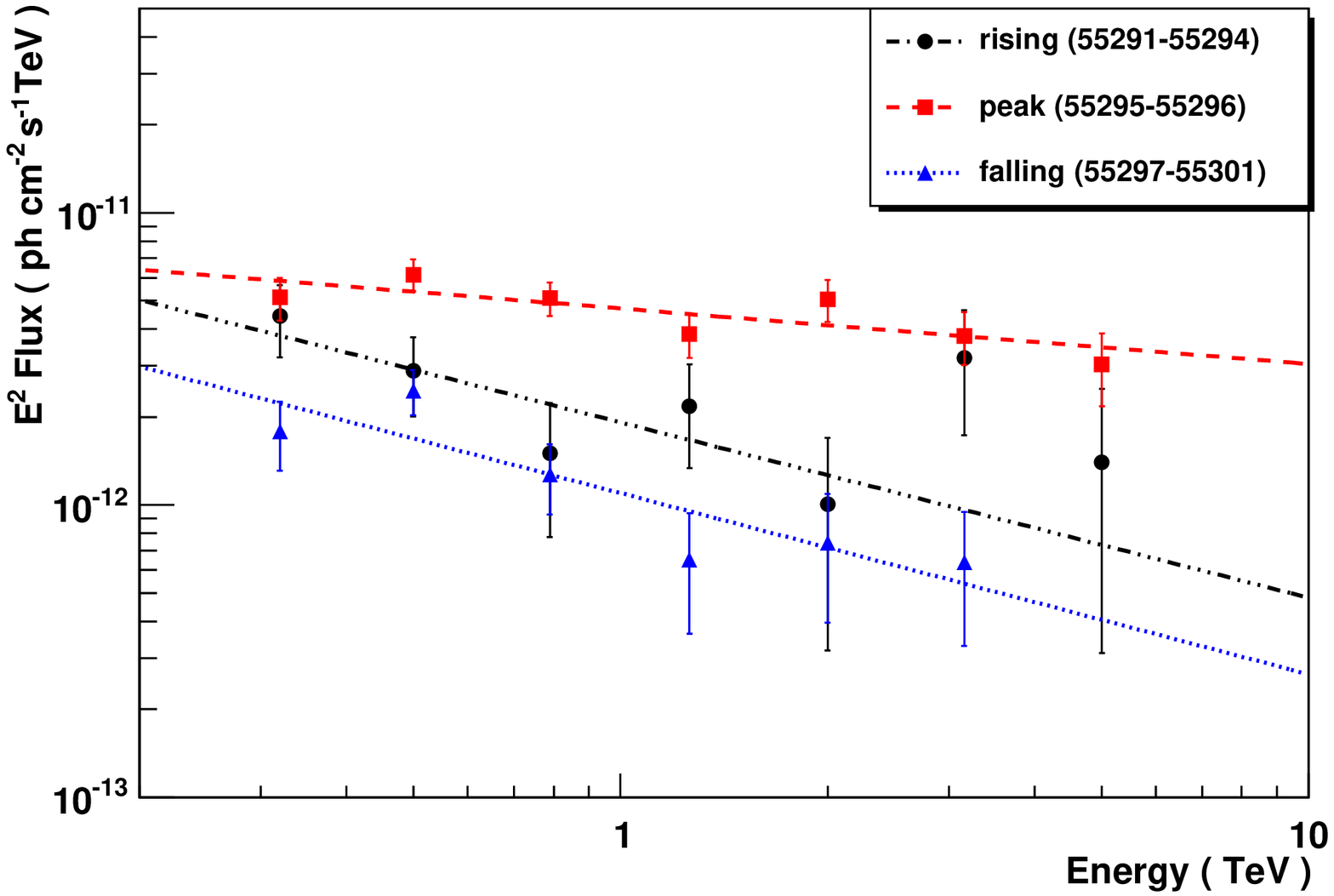}
\caption{Spectral measurements during three periods: leading up to the flare (MJD 55291--55294, April 5--8), peak of the flare (MJD 55295, 55296, April 9 and 10), and trailing the flare (MJD 55297--55301, April 11--15).  The lines are power-law fits to the data, with the values for the flux normalization constant and the spectral index given in Table \ref{fluxindextbl}.}
\label{spectra}
\end{figure}

Figure \ref{spectra} shows the spectra measured during the rising
period between April 5 and 8 (MJD 55291--55294), during the peak on
April 9 and 10 (MJD 55295--55296), and during the falling period
between April 11 and 15 (MJD 55297--55301).  
Power-law fits of the form $\Phi = \Phi_0 (E/\textrm{TeV})^{-\Gamma}$ are applied
to all three periods, and the corresponding power-law fit parameters are
listed in Table \ref{fluxindextbl}.  The spectral index of the peak
period differs from that of the falling period by $2.2\,\sigma$, and from that of the rising period by $1.3\,\sigma$.  The peak period has the hardest spectrum of all three periods.

\begin{deluxetable}{ccccccc}
\tablewidth{0pt}
\tablecaption{Spectral power-law fit parameters and hardness ratios for the three periods of the M\,87 flare in 2010 April: Rising, Peak, and Falling.  Errors given are statistical only.}
\tablehead{\colhead{Periods} & \colhead{MJD Date} & \colhead{Flux
    Normalization} & \colhead{Spectral Index} & \colhead{$\chi^2$/dof} & \colhead{Hardness Ratio} \\
 \colhead{} & \colhead{} & \colhead{Constant $\Phi_0$} & \colhead{$\Gamma$} & \colhead{}  & \colhead{} \\
 \colhead{} & \colhead{} & \colhead{$10^{-12}$
   ($\rm{cm}^{-2}\rm{\,s}^{-1}\rm{\,TeV}^{-1}$)} & \colhead{} & \colhead{} & \colhead{} \label{fluxindextbl} } 
\startdata
Rising & 55291--55294 & $1.92 \pm 0.42$ & $2.60 \pm 0.31$ & 4.1/4 & $0.35 \pm 0.12$\\
Peak & 55295--55296 & $4.71 \pm 0.29$ & $2.19 \pm 0.07$ & 4.3/5 & $0.28 \pm 0.03$\\
Falling & 55297--55301 & $1.10 \pm 0.16$ & $2.62 \pm 0.18$ & 5.2/4 & $0.10 \pm 0.04$\\
\enddata
\end{deluxetable}

A hardness ratio (HR) test is also applied to investigate further the possibility of spectral variability between these three different periods.  The HR may provide more sensitivity as it is obtained in a straightforward way from the energy distribution of the excess events, whereas the spectral index calculation requires multiple binning and fitting of the data.  The HR used here is defined as the ratio of the integral flux in the energy range 1--10\,TeV to that in the range 0.35--1\,TeV.  For the rising period, HR = $0.35 \pm 0.12$; for the peak period, HR = $0.28 \pm 0.03$; and for the falling period, HR = $0.10 \pm 0.04$ (see Table \ref{fluxindextbl}).  The HR for the peak period is found to be larger than that for the falling period with a statistical significance of $3.6\,\sigma$, compared to the $2.2\,\sigma$ for the spectral index difference of the same time periods.  The increased significance may result from a higher sensitivity of the HR to spectral variability.  However, we also note that the HR for the rising period is $2.0\,\sigma$ larger than that for the falling period while the spectral indices from these periods are identical.  This may be a result of statistical fluctuation due to the poor statistics of the rising period spectral measurements.

\begin{figure}
\plotone{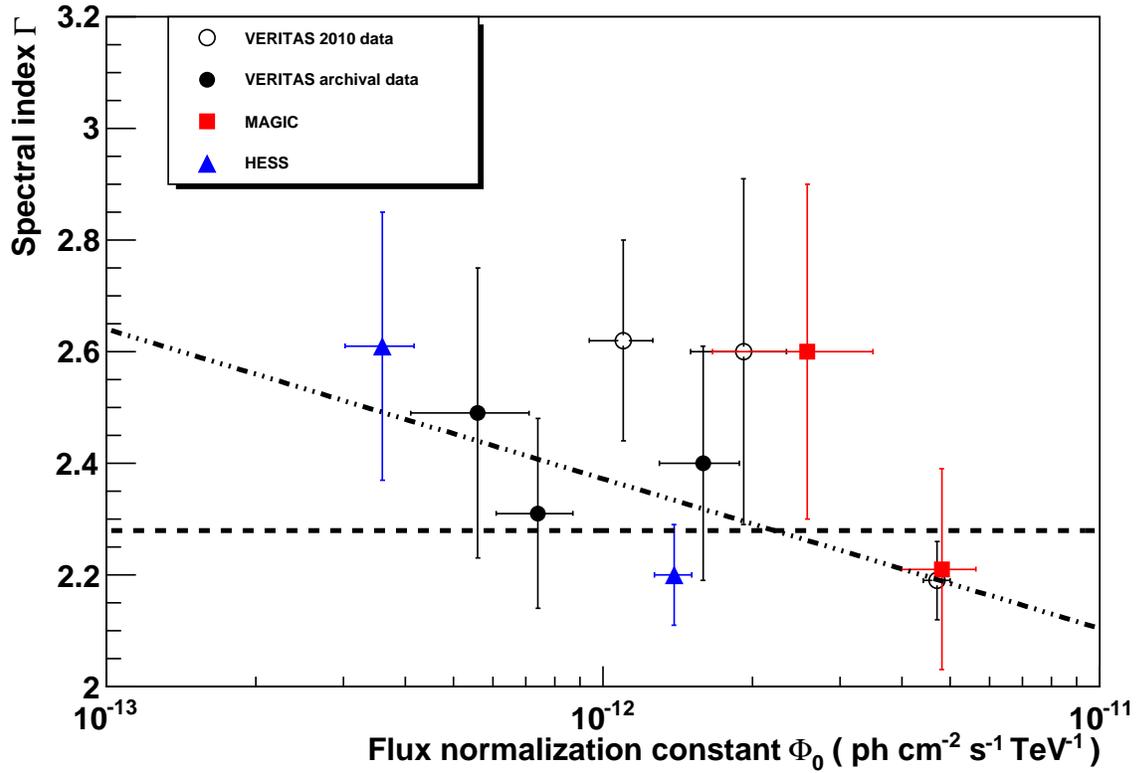}
\caption{Spectral index vs. flux normalization constant using spectra
  from the three periods (rising, peak, and falling) and archival spectra from
  2004 onward.  The dashed line represents a constant fit with a $\chi^2$
  probability of 0.26, and the dash-dotted line represents a linear fit of the
  form $\Gamma = p_0 + p_1 \mathrm{log}_{10}\Phi_0$ with a $\chi^2$ probability of 0.52.  The values of $p_0$ and $p_1$ are given in the text.}
\label{fluxindex}
\end{figure}

Figure \ref{fluxindex} shows the spectral index ($\Gamma$) plotted against flux
normalization constant ($\Phi_0$) for the 2010 April flare spectra
(open circles), together with archival VHE gamma-ray spectra from 2004
onward \citep{hess06,veritas08,magic08,joint,veri10}.  A constant-flux fit
to the 2010 April flare flux-index data yields a $\chi^2$ probability
of 0.05.  A linear fit of the form $\Gamma=p_0 +
p_1\mathrm{log}_{10}\Phi_0$ yields a $\chi^2$ probability of 0.67,
with the parameter $p_1 = -0.72 \pm 0.30$, which is $2.4\,\sigma$ away
from zero.  Although the fit may suggest a possible correlation
between the spectral index and the flux normalization constant, the
data do not provide definitive evidence for spectral variability
during this flaring episode.  Using all the
flux-index data available since 2004, a constant-flux fit yields a
$\chi^2$ probability of 0.26, while a linear fit yields a $\chi^2$
probability of 0.52 with $p_1 = -0.27 \pm 0.13$.

\section{Discussion}
VERITAS first detected M\,87 in 2007 in a low state, with
emission at $\sim 2$\% of the Crab Nebula flux above 250\,GeV \citep{veritas08}.  In 2008, VERITAS detected flaring activity up to 10\% of the
Crab Nebula flux above 250\,GeV during a joint monitoring campaign in which
correlations between VHE gamma rays, X-ray, and radio \citep{joint}
were found.  In 2009, M\,87 was observed to be in a low state
again at $\sim 1\%$ of the Crab Nebula flux above 250\,GeV \citep{veri10}.  In 2010 April, VERITAS observed the brightest emission ever seen from M\,87, with a flux up to 20\% of the Crab Nebula flux above 350\,GeV.  In comparison to previous constraints from past flares
\citep{hess06,magic08,veri10}, the 2010 VERITAS data set yields the
fastest exponential flux-changing time ($0.90^{+0.22}_{-0.15}$\,days) ever observed for M\,87.  This time constraint gives a new upper
limit on the emission region size that is lower than those derived
from previous observations.  Using the exponential decay time, the
emission region size has radius $R
\le R_{var} = \delta \,c \,\Delta t = 2.3 \times 10^{15}\,\delta$\,cm
$\approx 1.3 \delta\,R_s$, where $R_s$ is the Schwarzschild radius of the
M\,87 black hole ($= 2GM_{BH}/c^2 \approx 1.8 \times 10^{15}$\,cm, with $M_{BH} = 6.2 \times 10^9 M_{\odot}$ scaled from \citet{gebhardt11} to the distance used in this paper) and $\delta$ is the relativistic Doppler factor.  As in earlier findings, this may point to the black hole vicinity as the actual origin of the VHE radiation.  While an increased X-ray flux from the nucleus seems to support this hypothesis, no increase of the radio flux from the nucleus could be found \citep{joint2010}, in contrast to the contemporaneous radio and VHE gamma-ray flares observed in 2008 \citep{joint}.   
 
Another notable characteristic of the 2010 flare is the large
difference between the rise time and the decay time of the flux, a
feature which has not been seen in previous flares.  Since previous VHE flares in 2005 and 2008 were not sampled at a comparable accuracy and their onsets were not as well defined as the 2010 flare, this is the first M87 VHE flare that allows the determination of the rise and fall times.  The shape of the 2010 flare also seems less erratic as compared to the earlier flares, which could point to a different production mechanism.  However, given the lower statistics of the earlier flares, this is difficult to quantify and requires future observations to disentangle.

From a compilation of multiwavelength data sets spanning decades,
\citet{jointfermi} presented a spectral energy distribution (SED) of
M\,87, along with hadronic and leptonic models.  The hadronic
synchrotron-proton blazar (SPB) model \citep{reimer04} suggests gamma-ray
emission from synchrotron radiation by protons or by muons and pions.
However, the SPB model SED produced using archival data before 2004 shows a steep drop-off at TeV energies that is not compatible with the spectra
obtained from the 2010 data set or with any previous VHE spectral measurements.  \citet{barkov2010} proposed a scenario where a red giant star, with an envelope of loosely bound material, interacts with the base of the jet.  VHE gamma rays are produced near the supermassive black hole via proton--proton interactions between the jet and the red giant cloud.  The gamma-ray light curve produced from this model shows an exponential increase/decay time of $\sim 1$\,day, identical to the decay timescale obtained from the 2010 data.  However, the model gave no prediction on the VHE spectrum for comparison with our data.

There are also several leptonic jet models with different geometric structures that can explain the VHE gamma-ray emission, such as the decelerating jet model by \citet{georg05}, the multi-blob synchrotron-self-Compton (SSC) model by \citet{lenain08}, and the spine-sheath model by \citet{tav08}.  The SED solutions obtained from these models can explain the observed rapid variability and match the VHE spectra well in both low and flaring states from 2005 and earlier.  Looking into some of these models in more detail, the model parameters can be adjusted to account for recent measurements.  In the case of the multi-blob scenario, \citet{lenain08} showed that for the case of M\,87, the model spectrum hardens with decreasing magnetic field.  In order to keep the size of the VHE emitting region of the order of the Schwarzschild radius, the local value of the magnetic field should be $\leq 0.01$\,G.  The spine-sheath model, however, seems to face difficulties in achieving a harder spectrum due to absorption of TeV photons from interactions with the optical--IR photons from the spine.  As pointed out by \citet{tav08}, severe gamma-ray photon absorption can be alleviated by increasing the emission region size, which would decrease the absorption optical depth.  However, this would be limited by the observed short-term variability.  \citet{fermi09} fitted a homogeneous one-zone SSC model using 2009 VLBA radio, \emph{Chandra} X-ray, and \textit{Fermi}-LAT measurements when M\,87 was observed to be in a low state from radio to VHE gamma rays.  A contemporaneous spectral measurement in the VHE range was not possible due to low statistical significance \citep{veri10}, but compared to archival low-state VHE measurements, the one-zone SSC model seems to underestimate the VHE gamma-ray flux by almost an order of magnitude.  \citet{georg05} and \citet{lenain08} demonstrated that one-zone homogeneous models are unlikely to reproduce the observed VHE spectrum.

\citet{giannios2010} presented a scenario where minijets are formed within the jet due to flow instabilities.  These minijets move relativistically with respect to the main jet flow.  VHE gamma rays are produced from the interactions between the minijets and the jet, and are beamed with large Doppler factor when the minijets are aligned with our line of sight.  The minijets model SED is compatible with the 2010 data.  A satisfactory solution for the high state observed by VERITAS in 2010 is also possible within the magnetosphere model \citep[e.g.,][]{nerov07,Rieger08,vincent10,levin11}.  The magnetosphere model is dependent on the injected plasma which suggests that a vacuum gap with a large electric field that is capable of accelerating electrons to very high energies may be formed during a period of low accretion rate.

We cannot discriminate between different leptonic models based on this VHE data alone.  The spectral change with flux level would serve as an important input for the modeling once it is confirmed by a second flare.  Leptonic models tend to predict a more direct correlation between X-ray and VHE gamma rays.  For the 2010 flare of M\,87, extensive follow-up observations of the VHE gamma-ray flare \citep{atel2542} were
carried out in X-ray, optical, and radio wavebands.  The result is a
much more complete sampling across different energy bands than in the
case of previous M\,87 flares, providing a data set that will help to
constrain the emission region and the radiative processes involved.  A
separate, upcoming publication \citep{joint2010} will present the multiwavelength result, which spans 16 decades of energy.

\acknowledgments
This research is supported by grants from the U.S. Department of Energy Office of Science, the U.S. National Science Foundation and the Smithsonian Institution, by NSERC in Canada, by Science Foundation Ireland (SFI 10/RFP/AST2748) and by STFC in the U.K. We acknowledge the excellent work of the technical support staff at the Fred Lawrence Whipple Observatory and at the collaborating institutions in the construction and operation of the instrument.

{\it Facility:} \facility{VERITAS} 

\bibliographystyle{apj}
\bibliography{ref}

\end{document}